\documentclass[12pt]{article}
\usepackage{jeosman}
\usepackage{epsfig}
\usepackage{amsmath}
\usepackage{amssymb}

\usepackage{mathrsfs}

\newcommand{\rrr}{\mathbf{r}}
\newcommand{\EEE}{\mathbf{E}}
\newcommand{\nablabf}{\mathbf{\nabla}}

\begin{document}

\title{Nanostructure design for surface-enhanced Raman spectroscopy -- prospects and limits}

\author{Sanshui Xiao,$^{*\ddag}$ Niels Asger Mortensen,$^{*\ddag}$\footnote{email: asger@mailaps.org} Antti-Pekka
Jauho$^{\ddag\dag}$}

\address{$^*$Department of Photonics Engineering, Technical
University of Denmark, DTU Fotonik, DTU-building 345 west, DK-2800
Kongens Lyngby\\
$^\ddag$Department of Micro and Nanotechnology, Technical University
of Denmark, DTU Nanotech, DTU-building 345 east, DK-2800 Kongens
Lyngby, Denmark\\$^\dag$Laboratory of Physics, Helsinki University
of Technology, P. O. Box 1100, FI-02015 HUT, Finland}



\begin{abstract}
Surface-enhanced Raman spectroscopy (SERS) allows single-molecule
detection due to the strong field localization occurring at sharp
bends or kinks of the metal-vacuum interface.  An important question
concerns the limits of the signal enhancement that can be achieved
via a judicious design of the surface.  By using a specific example
of a technologically realizable nanopatterned surface, we
demonstrate that while very high enhancement factors ($\approx
10^{12})$ can be found for an ideal surface, these are unlikely to
be achieved in laboratory samples, because even a minute, inevitable
rounding-off strongly suppresses the enhancement, as well as shifts
the optimal frequency.  Our simulations indicate that the geometric
enhancement factors are unlikely to exceed $\approx 10^8$ for real
samples, and that it is necessary to consider the geometric
uncertainty to reliably predict the frequency for maximum
enhancement.
\end{abstract}

\maketitle

\section{Introduction}

\begin{figure}
\begin{center}
\epsfig{file=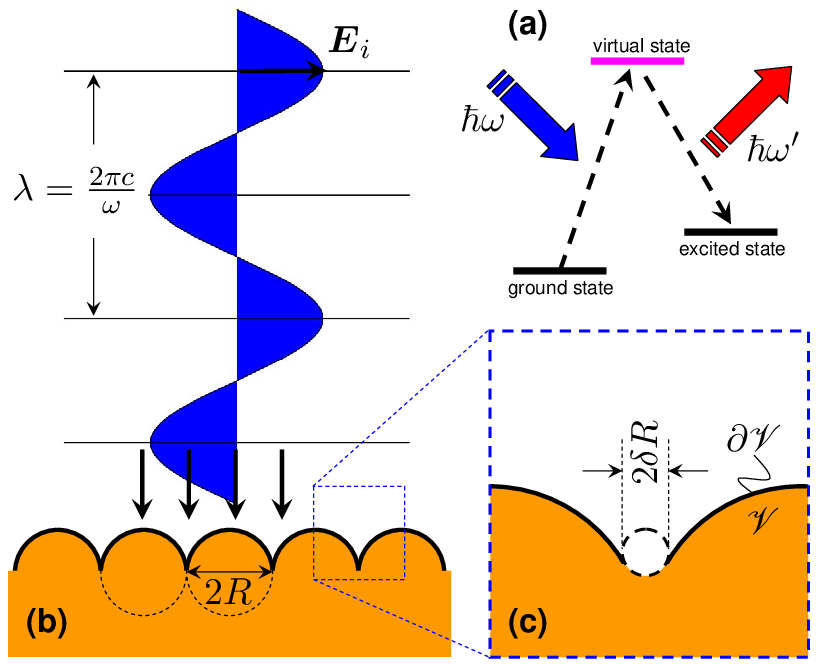, width=0.7\columnwidth,clip}
\end{center}
\caption{(color online) (a) Energy diagram for the Raman
scattering process involving the absorption of an incident photon
with energy $\hbar\omega$, and a subsequent emission of a photon
with energy $\hbar\omega'$ via an intermediate virtual state in a
molecule. The scattering is inelastic and the scattered light is
Stokes shifted by $\Delta\omega=\omega-\omega'$. (b) Metallic
periodic structure composed of infinite half-cylinders of radius
$R$. The normal incident electrical field $\EEE_i$ is polarized
transverse to the cylinders. (c) Magnification of the groove
structure indicating the nanoscale cutoff modelled by a finite
radius of curvature $\delta R$.} \label{fig1}
\end{figure}

Raman spectroscopy is a powerful technique which has become one of
the optical workhorses in analytical chemistry where it is used to
detect and quantify a variety of chemical substances including
biomolecules absorbed to a substrate. The 1974 report by Fleischmann
\emph{et al.}~\cite{Fleischmann:1974} on surface-enhanced Raman
spectroscopy (SERS) created an avalanche in surface-enhanced
spectroscopies~\cite{Moskovits:1985}. Localized surface plasmons may
dramatically enhance the otherwise intrinsically weak Raman
scattering [see panel (a) in Fig.~\ref{fig1}] and this has paved the
way for chemical detection in very dilute solutions, reaching the
limit of single-molecule detection~\cite{Kneipp:1997}. Initially,
there was a strong emphasis on disordered and rough metal
surfaces~\cite{GarciaVidal:1996}; however during the past decade the
use of highly engineered surface plasmonic structures has been
envisioned~\cite{Pendry:2004a}. The latter development has to a
large extent been fueled by the rapid development of a variety of
nanofabrication technologies, including high-resolution
electron-beam lithography on the true nano scale. As a consequence,
surface plasmon subwavelength optics has in general developed
tremendously~\cite{Barnes:2003}, and in particular SERS has been
observed for a large variety of nanostructures, such as nanorods,
nanorings, triangular nanoprisms, nanocubes, nanocrescents, and
nanorice~\cite{Lal:2007}.

Field localization is the key to understanding and further enhancing
SERS. Structures with pronounced bends in their surface
$\partial{\mathscr V}$ will tend to localize the field also in the
plane of the surface~\cite{Moreno:2006}. Singular points, such as
sharp kinks and corners, may lead to nonanalytic or even
mathematically divergent electric fields. Real structures are
smooth, of course, at least on some microscopic length scale. All
fabrication processes have their inherent upper bounds on the
sharpness that can be fabricated reproducibly. As an example,
state-of-the-art groove waveguides have sharp features down to a
10~nm scale~\cite{Bozhevolnyi:2007}. In an ensemble of
nanostructures there will always be a size dispersion or
structure-to-structure fluctuations in some details of the geometry,
for example the nanoscale rounding $\delta R$~\cite{roundning} will
differ from structure to structure, with possibly far-reaching
effects on the SERS. Fluctuations between repeated features and
variations in the smoothness of individual features have previously
been addressed for fractal structures in a number of papers by
S{\'a}nchez-Gil \emph{et al.}, see e.g.
Refs.~\cite{SanchezGil:2000,SanchezGil:2002}. The pronounced
sample-to-sample fluctuations are potentially a serious challenge to
the development of a technology for reliable use in quantitative
sensing applications and bio-chemical analysis.

In this paper we address another crucial issue: how are the high
enhancement factors found for highly ideal surface structures
affected by round-off effects?  By means of numerical examples we
shall illustrate the extreme precision needed in the fabrication of
nanostructured surfaces, if one aims at SERS in the excess of ten
orders of magnitude. Specifically, we consider the simple model
geometry in panel (b) of Fig.~\ref{fig1}. Historically, this was the
first structure considered theoretically to explain the large
enhancement occurring for rough metal
surfaces~\cite{GarciaVidal:1996}. Today, we may easily envision
realization of groove structures~\cite{Bozhevolnyi:2006} closely
resembling the geometry considered in panel (c), though
state-of-the-art structures may still be somewhat larger in scale.
As mentioned above, the points of intersection of neighboring
cylinders may lead to a singularity in the electric field, and
consequently arbitrarily large enhancement factors [see
Eq.~(\ref{eq:gamma}) below].

\section{Theory}

In the context of SERS it is customary to estimate the
surface-enhancement factor $\gamma$ from~\cite{GarciaVidal:1996}
\begin{equation}\label{eq:gamma}
\gamma(\omega)\equiv\frac{\int_{\partial{\mathscr
V}}d\rrr\,\gamma(\rrr,\omega)}{\int_{\partial{\mathscr
V}}d\rrr},\; \gamma(\rrr,\omega)\approx
\frac{\big|\EEE_t(\rrr,\omega)\big|^4}{\big|\EEE_i(\rrr,\omega)\big|^4},
\end{equation}
where $\rrr$ is the position of the molecule. To a first
approximation the surface-enhancement factor is independent on the
intrinsic properties of the molecule.  Here, $\partial{\mathscr V}$
is the surface bounding the volume ${\mathscr V}$ occupied by metal,
$\EEE_i$ is the incident field, and the total field is
$\EEE_t=\EEE_i+\EEE_s$ with $\EEE_s$ being the scattered component
that arises in the presence of the metal surface. The
surface-averaged enhancement factor applies to the case of a
spatially uniform surface-coverage of molecules.

Our theoretical study takes the wave equation for the electrical
field as a starting point
\begin{equation}\label{eq:wave}
\nablabf\times\nablabf\times
\EEE(\rrr)=\frac{\omega^2}{c^2}\varepsilon(\rrr,\omega)\EEE(\rrr)
\end{equation}
where $\omega$ is the angular frequency and $c$ is the vacuum
velocity of light. For structures as in Fig.~\ref{fig1} the
spatially dependent relative dielectric function is of the form
\begin{equation}\label{eq:eps}
\varepsilon(\rrr,\omega)=\left\{\begin{array}{ccc}1 &,&
\rrr \notin {\mathscr V},\\\\
\varepsilon(\omega) &,& \rrr\in {\mathscr V},
\end{array} \right.
\end{equation}
where $\varepsilon(\omega)$ is the dielectric function of the
metal. To make our simulations mimic a real material we use the
bulk dielectric constant of silver (thus neglecting spatial
dispersion near the metal surface), illustrated in Fig.~\ref{fig2}
which shows standard handbook data for $\varepsilon(\omega)$
~\cite{Palik:1985} along with a numerical fit to the Drude
expression
\begin{equation}\label{eq:Drude}
\varepsilon(\omega)=
\varepsilon_\infty+\frac{\left(\varepsilon_0-\varepsilon_\infty\right)\omega_p^2}{\omega^2+
i\omega/\tau}
\end{equation}
with $\omega_p$ being the plasma frequency and $\tau^{-1}$ the
damping rate. This simple model accounts very well for the real
part, however some deviations are seen in the visible for the
imaginary part. The spectral position of resonances will thus be
accurately predicted by using the Drude fit while we shall see that
even small deviations in the imaginary part of $\varepsilon$ may
influence the enhancement factor, Eq.~(\ref{eq:gamma}), by orders of
magnitude.

\section{Numerical results}

In the following we use a finite-element approach (Comsol
Multiphysics) to solve Eq.~(\ref{eq:wave}) in a unit cell resembling
panel (c) in Fig.~\ref{fig1}. We apply periodic boundary conditions
in the direction parallel to the surface. Deep inside the metal and
far out in the free space we employ perfectly-matched layers in the
perpendicular direction~\cite{Berenger:1994}, so that the domain has
an extension $\gg \lambda$, i.e. significantly exceeding the
free-space wavelength. Eq.~(\ref{eq:wave}) is solved for a normal
incident electrical field polarized perpendicular to the groove
axis, see panel (b) in Fig.~\ref{fig1}, which strongly polarizes the
surface by excitation of a surface-plasmon resonance. With the
resulting scattered field, and hence $\EEE_t$, we subsequently
evaluate Eq.~(\ref{eq:gamma}) by a boundary integration routine. Our
meshing is  chosen to ensure convergence of the enhancement factor;
this requires a particularly fine mesh in the groove region. As an
example, resolving the peak in $\gamma$ at resonance has typically
required a sub-nanometer meshing. On a standard-equipped PC the
calculation of a single near-resonance frequency point typically
takes around 30 seconds.

\begin{figure}
\begin{center}
\epsfig{file=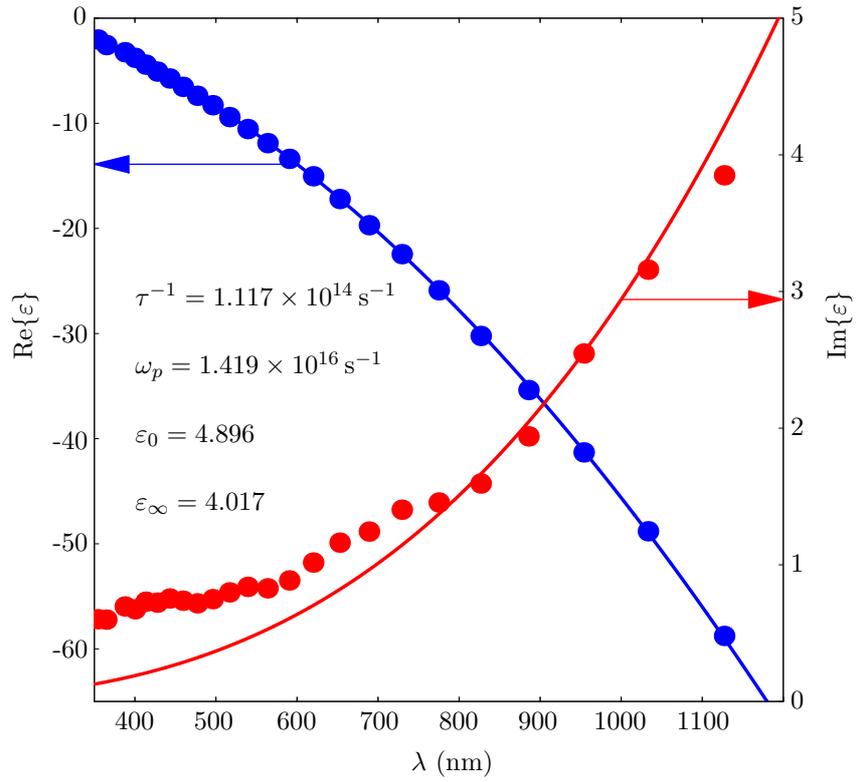, width=0.7\columnwidth,clip}
\end{center}
\caption{(color online) Dielectric function of silver. The data
points show handbook data from Ref.~\cite{Palik:1985} while the
solid lines are fits to the Drude model in Eq.~(\ref{eq:Drude}).}
\label{fig2}
\end{figure}

\begin{figure}
\begin{center}
\epsfig{file=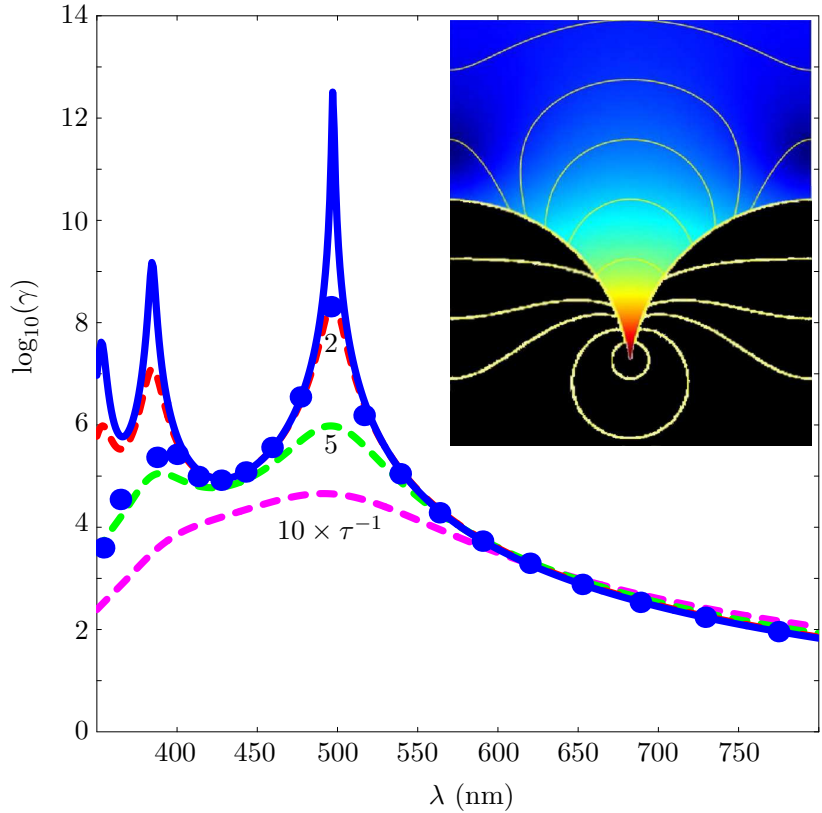, width=0.7\columnwidth,clip}
\end{center}
\caption{(color online) Wavelength dependence of the
surface-averaged enhancement factor for a structure with $R=15$~nm
and $\delta R=0.1$~nm. The data points result from direct
simulations at discrete frequencies using handbook data for
silver, see Fig.~\ref{fig2}, while the solid line shows the
corresponding results based on the Drude model,
Eq.~(\ref{eq:Drude}). The dashed lines illustrate the suppression
of the SERS effect for increasing damping rates $\tau^{-1}$. The
inset shows $\gamma$ at the main peak where the electrical field
is strongly localized in the groove. The electrical field points
in a direction parallel to the superimposed electric field lines.}
\label{fig3}
\end{figure}

Figure~\ref{fig3} shows the wavelength dependence of the
surface-averaged enhancement factor in the case of $R=15$~nm and
$\delta R=0.1$~nm. The solid line shows results based on the Drude
model, Eq.~(\ref{eq:Drude}), while the points are obtained from a
simulation using the discrete frequencies where handbook data is
available, see Fig.~\ref{fig2}. First, we note that these
geometrical parameters easily lead to surface-averaged enhancement
factors in excess of $10^{12}$, i.e. twelve orders of magnitude,
in the visible. Second, when inspecting the peak around
$\lambda\sim 500$~nm and comparing the two simulations, it is also
clear that the Drude model, Eq.~(\ref{eq:Drude}), overestimates
the enhancement by perhaps up to three orders of magnitude. This
overestimation may be traced back to the imaginary part of
$\varepsilon$ shown in Fig.~\ref{fig2} where the Drude model
seriously underestimates the broadening of the surface-plasmon
resonance. Finally, when approaching the ultra-violet, resonances
predicted by the Drude model are barely visible when employing the
handbook data directly. While general trends can be extracted from
a Drude description, these results emphasize the need for an
accurate material-dispersion law for even semi-quantitative
predictions of SERS. In particular, the surface-plasmon damping is
crucial in the vicinity of resonances as clearly seen in the
dashed traces where we for illustrative purposes have increased
the damping rate $\tau^{-1}$ by a factor of two, five, and ten.

\begin{figure}
\begin{center}
\epsfig{file=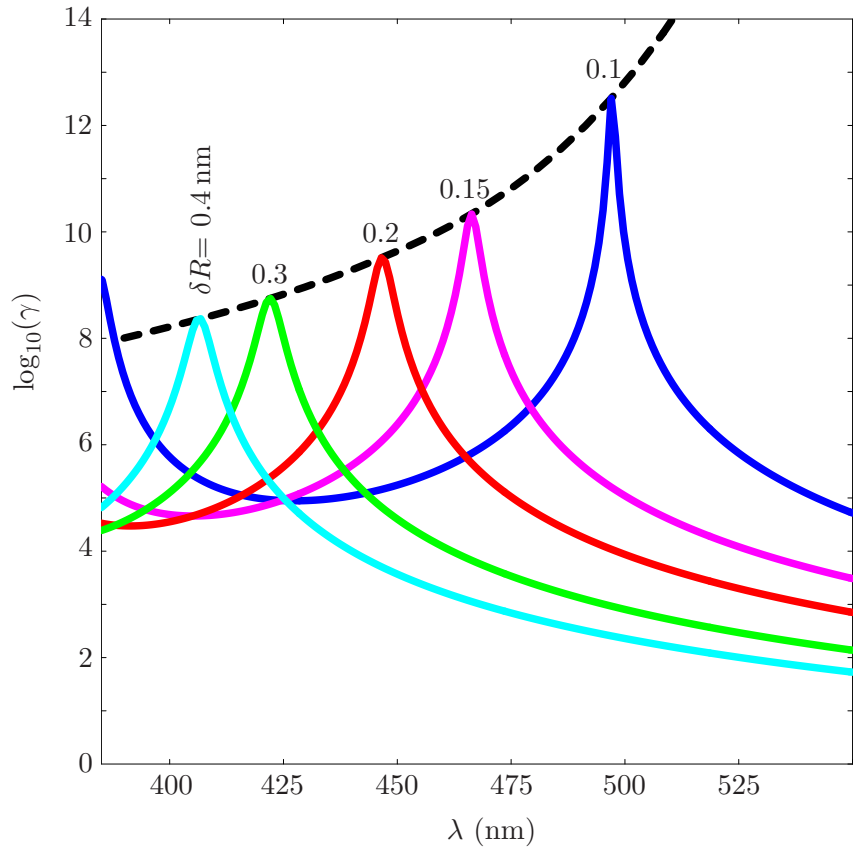, width=0.7\columnwidth,clip}
\end{center}
\caption{(color online) Wavelength dependence of the
surface-averaged enhancement factor for a structure with $R=15$~nm
and varying nanoscale cutoff $\delta R$. } \label{fig4}
\end{figure}

While previous studies~\cite{GarciaVidal:1996} have reported a
spectrally broadened enhancement, our detailed finite-element
simulations suggest the existence of highly peaked enhancement
factors associated with the formation of a surface-plasmon resonance
strongly localized in the groove. A related transverse localization
has recently been reported experimentally for groove
waveguides~\cite{Bozhevolnyi:2006} and subsequently confirmed
numerically~\cite{Moreno:2006}. Obviously, spatial localization is
the route to high enhancement and especially one should minimize the
overlap with the lossy metal causing Ohmic broadening of resonances.
The nanoscale cutoff $\delta R$ is the geometrical knob that
controls the strength of localization and the overlap with the
metal. A stronger localization ($\delta R \rightarrow 0$) will
usually be expected to squeeze up the resonance in frequency space.
However, when increasing $\delta R$ we actually observe a blue-shift
which can be understood within the quasi-static framework of
Ref.~\cite{Wang:2006} [see their Eq.~(4)] where an increase in metal
volume ${\mathscr V}$ causes modes at $\omega$ to blue-shift to a
frequency $\omega+\Delta\omega$ with a less negative dielectric
function, i.e. $\varepsilon(\omega+\Delta\omega)>
\varepsilon(\omega)$. In Fig.~\ref{fig4} we illustrate this point
for varying values of $\delta R$ at fixed $R=15$~nm. For simplicity
the results are based on the Drude model. Our results show that even
for a relatively conservative value of the ratio $\delta R/R\sim
3\%$ the enhancement has already dropped to $\gamma\sim 10^8$
compared to $\gamma\sim 10^{12}$ in the case of $\delta R/R\lesssim
1\%$.

\begin{figure}
\begin{center}
\epsfig{file=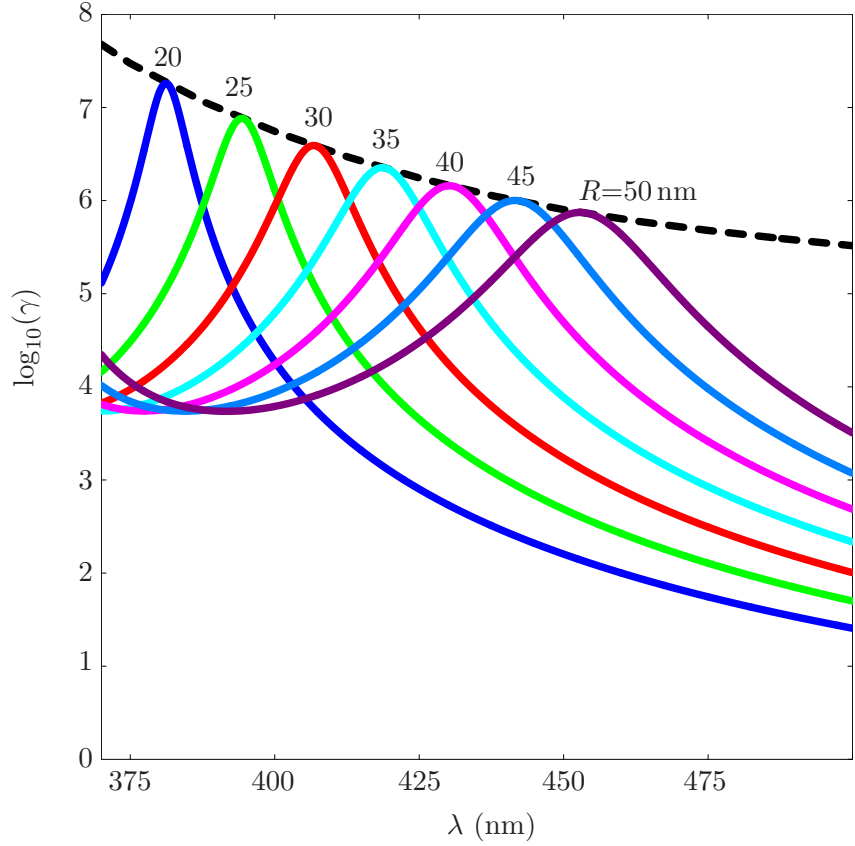, width=0.7\columnwidth,clip}
\end{center}
\caption{(color online) Wavelength dependence of the
surface-averaged enhancement factor for a structures with varying
radius $R$ of the semi cylinders and a nanoscale cutoff $\delta
R=1$\,nm. } \label{fig5}
\end{figure}

Another geometric factor affecting the  field enhancement is the
radius of cylinders. In Fig.~\ref{fig5} we illustrate this point
for varying values of $R$ at fixed $\delta R=1$~nm, again using
the Drude model. Our results show a red-shift of the resonance
when increasing the radius while the magnitude decreases slightly.
The red-shift may be understood qualitatively from the scale
invariance of Eq.~(\ref{eq:wave}) in the absence of  material
dispersion. Figs.~\ref{fig4}~and~\ref{fig5} together illustrate
how the enhancement is influenced by both length scales $R$ and
$\delta R$ in a nontrivial interplay.

\section{Conclusion}

In conclusion, we have presented finite-element simulations of the
SERS enhancement factor for an example of a nanostructured surface,
which represents the state-of-the-art technology. We find that a
well-defined structure, with mathematically sharp boundaries [see
Eq.~(\ref{eq:eps})], may exhibit huge enhancement factors,
$\gamma\approx 10^{12}$. But any geometric smoothening rapidly
decreases the enhancement, and even though a computational
optimization may be possible by a painstaking survey of the
parameter space ({\it given} that the dielectric function of the
material is known very accurately - no small feat by itself!), it is
doubtful whether the extremely high theoretical enhancement factors
can be reached in practice and in a reproducible manner.

\section*{Acknowledgments}

We acknowledge stimulating discussions with S.~I. Bozhevolnyi. This
work is financially supported by the \emph{Danish Council for
Strategic Research} through the \emph{Strategic Program for Young
Researchers} (grant no: 2117-05-0037) and the \emph{Danish Research
Council for Technology and Production Sciences} (grant no:
274-07-0379) as well as the FiDiPro program of the Finnish Academy.

\newpage


\newpage

\end{document}